# The NC State All-campus Data Science and AI Project-based Teaching and Learning (ADAPT) Model: A mechanism for interdisciplinary engagement in workforce-relevant learning.

Dr. Rachel Levy, Executive Director of the Data Science and AI Academy and Professor of Mathematics, North Carolina State University, Dr. James B. Harr III, Assistant Teaching Professor of Data Science at the College of William and Mary, and Dr. David Stokes, Director of Data Science and AI Academic Programs, NC State University Data Science and AI Academy

Last Updated April 1, 2026

## Abstract

Academic institutions have been challenged to adapt as data science and AI have rapidly evolved into disciplines, degrees and careers. Efforts to provide students with learning experiences have led to the development of novel credentials, renamed departments, new schools and even additional colleges within universities. Generally, these approaches are siloed in some way - perhaps separating STEM students from those in the humanities or separating faculty assigned to these courses from their colleagues in their home departments. NC State University decided to take a novel approach by creating a new type of entity called an Academy that would reach across all disciplines, departments, colleges, centers and institutes to catalyze work in data science and AI in all points of the university's mission: teaching, research and engagement.

## Introduction

The **A**ll-campus **D**ata Science and **A**I **P**roject-based **T**eaching and learning model (known as the ADAPT model) was developed in 2021 by the Data Science and AI Academy at NC State University to guide instruction and build coherence among a wide variety of courses for students of all majors. The novel ADAPT educational model has attracted learners across the 12 colleges of the university, provided professional development for industry and government and guided teacher and student learning for K-12 settings. In only five years, the university has combined these new opportunities with existing programmatic strengths to enhance degrees and build new credentials.

Here we review and reflect on the first five years of the Data Science and AI Academy and document the development of the ADAPT model, which was designed to be replicable and support teaching and learning of data science and AI across the disciplines. This includes an introduction of the model, a description of courses that have put the model into practice, the staffing model for the courses, credentials that incorporate the courses, data describing the growth in the number of course sections, the instructor professional learning community, learner supports such as undergraduate course collaboration leaders, related research grants and conferences, and extensions to educational spaces beyond the university.

We present preliminary evidence that this model can inspire other institutions to adopt aspects of this approach and apply them in their own context. We hope this paper will provide a foundation for further collaboration, idea exchange and application of our successful ADAPT model.

## A New Type of Academic Unit: the NC State Academy

Around 2010, the emergence of data science and artificial intelligence as new disciplines, careers and key technological developments led to new educational considerations. Academic institutions began to wrestle with how to include data science and artificial intelligence (AI) in curriculum – from basic literacy to advanced discovery. The rapid development of theory and technologies coupled with their early adoption by industry and later by the public necessitated an approach to workforce preparedness within higher education that could facilitate both rapid prototyping and continuous reimagining to keep pace with the ever-changing landscape of data science and AI.

By 2020 NC State University was developing a strategy to build on its interdisciplinary strengths and existing programs in areas such as business, statistics, mathematics, computer science, geospatial analytics, operations research, design and plant sciences, to name only a few. These programs were building interdisciplinary graduate degrees in data science and AI that complemented the long-running Institute for Advanced Analytics master's degree in applied data science, which was established in 2007 as one of the first professional master's degrees in data science. Yet, a dedicated data science and AI-focused interdisciplinary space to expand offerings to all NC State colleges did not exist.

Rather than set up a new major, department, school or institute, NC State took a different approach – a new concept called an Academy.  The word academy was not associated with any formal type of entity in the University of North Carolina System and connoted the interdisciplinary National Academies of Science Engineering and Medicine.  The new Data Science (and soon also Artificial Intelligence) Academy grew out of the Data Science Initiative of the Office of Research and Innovation with the promise that at NC State "Data Science is for Everyone." With the launch of the DSA in July 2021, the Provost office stated that

> *At NC State, an academy identifies a true university-wide effort involving all 10 colleges. It encompasses multiple departments, centers and institutes and addresses all three pillars of the university's land-grant mission: education, research and service to the state of North Carolina.*

This enabled the Data Science and AI Academy (known as DSA) to function as a "skunkworks" – a unit that could operate within the regulations of the institution, but outside existing norms and practices in ways that would be beneficial to the whole institution (thanks to entrepreneur Charles Gaddy for this comparison).  Compared to existing units, the scope of the DSA would be required to be unusually broad and reach all of campus – including students, staff, faculty, alumni, community members and campus partners.  The goal would be to mind and address the gaps by networking across campus, to notice what capacity was needed in data science and AI, and then to collaboratively develop new programs or enhance existing ones. Primary gaps included a lack of data science credential pathways available to all students, research enablement to support data science research across campus, programming that could network the community and foster interdisciplinary collaboration, and outreach to K-12 and community colleges to collaborate on data science and AI learning.  **Here we focus on the curricular component – a novel approach to building data science and AI courses and credentials that could serve learners at all stages, across campus and beyond.**

## One-credit courses: an idea that stuck

The original DSA concept included piloting six new 1-credit courses that have persisted as core topics: Introduction to R/Python for Data Science, Data Visualization, Exploratory Data Analysis, Introduction to Big Data, Natural Language Processing and Ethics of Data Science.  While many institutions were piloting general "intro to data science" courses in the traditional 3-credit structure, such as UC Berkeley's Data 8 (University of California Berkeley, 2023), and Purdue's Data Mine (Gundlach & Ward, 2021), the DSA took a different approach to the course structure to include the components and benefits below.

- **The 1-credit structure allows for application-focused exploration** and appeals to students who want to try something new or add data science into their schedule of required courses.
- Offering 1-credit courses at three levels **accommodates students with a variety of prior experience levels**. The 200 level courses have no prerequisites, the 400 level courses list skill-based prerequisites and the 500 level courses at the graduate level are more research-focused.

- By **listing prerequisite skills rather than courses, students can self-select** according to their prior experiences.
- The topical and interdisciplinary nature of the courses **appeals to students from across majors** in the university's 11 colleges and almost 90 departments.
- The courses can **fit into busy schedules** that vary widely across programs. Both students and faculty have been willing to add a 1-credit course as an overload if it didn't fit into their base schedule and
- Instructors from industry and government can teach courses on top of their full-time jobs.
- The mode of delivery can be **structured like a workshop** with multiple points of contact across the semester rather than a single intensive experience with the intention to promote retention.
- The limited number of meetings as a **design constraint focuses the courses** on digging into a topic and gaining experience rather than comprehensive overview or theoretical mastery.
- The 1-credit courses could **be incorporated into existing credentials** such as majors, concentrations, minors and certificates within other departments.
- As a **catalyst for new credentials**, the 1-credit courses can provide breadth and could be an entry point, supplement or complement to departmental 2 or 3-credit courses for depth.

The DSA strategy has been to recruit instructors from current NC State educators , industry partners , government and faculty from other academic institutions who are able to teach a wide variety of topics that incorporate multiple disciplinary perspectives and current workforce trends. In Spring 2026, the instructor group percentages were 42%, 19%, 12%, and 27% respectively. The ADAPT model, described below, was created to allow for this rich variety of perspectives and expertise while providing a consistent and coherent DSA courses learning experience. ,

## The ADAPT Model

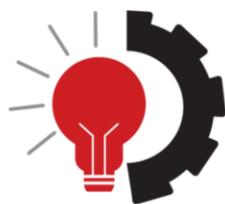
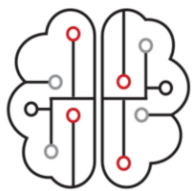
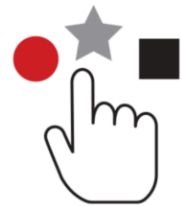

Project-based Learning | 10 Common Learning Elements | Workforce Preparedness

# ADAPT
All-campus Data science and AI Project-based Teaching and learning model

*Figure 1: Graphic identifier for the All-Campus Data Science and AI Project-Based Teaching and Learning Model, noting the three components described in the text.*

The ADAPT model is grounded in research related to pedagogical approaches such as project-based and problem-based learning and instructional design principles that prioritize higher levels of Bloom's taxonomy (Anderson & Krathwohl, 2001; Kokotsaki et al., 2016; Thomas, 2000). Figure 1 shows the three primary components of DSA's ADAPT model: project-based learning, 10 common learning elements and workforce preparedness. Note that the acronym ADAPT intentionally connotes that this is a living model that should change over time in response to research and feedback and can be adapted to new educational contexts. Even within DSA, the model has been successfully applied in non-credit-bearing courses for industry and graduate programs as well as courses and camps for students in K-12 and teacher professional development.

### ADAPT Model Component 1: Project-based Teaching and Learning

All DSA students engage in a project, often scaffolded by the instructor with checkpoints over the semester. As part of the project-based approach, DSA courses require instructors to employ formative and summative assessments based on performance and applications of the course concepts, through assignments, projects and presentations rather than quizzes, tests and exams. This approach aligns with the concept of a workshop-style experience and is relevant to the way professional work is produced and evaluated in the workplace.

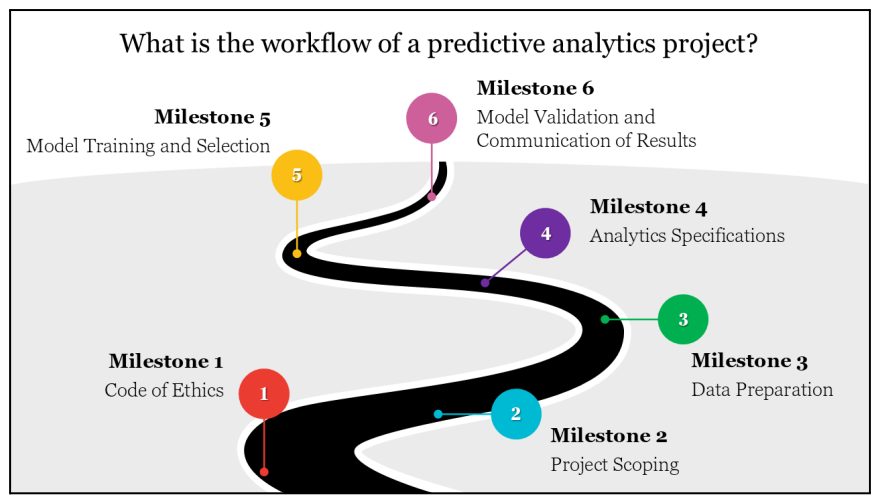

*Figure 2: Example of 7 milestones for projects in Predictive Analytics for Improving Services. (Credit: Kristin Porter)*

While another institution using the ADAPT model could choose a different method of evaluation based on the context of their program, this assignment and assessment policy helps reinforce the concept of project-based teaching and learning. Because the topics offered vary greatly, the assignments do as well, but they might include reflective writing, pseudo-code, coding or code commenting, online or in person discussions, reports, presentations and projects that tackle various elements of a data science workflow (Sample, M., May, 2011). Project development often occurs collaboratively within the professional learning community of instructors, who apply a backwards design approach in which they first think about what skills the students will demonstrate through the project and then scaffold the learning in the course to build up to that project (Wiggins & McTighe, 2005).

**ADAPT Model Component 2: Ten Common Learning Elements**

The ten common ADAPT learning elements guide instruction to include perspectives, practices and discoveries relevant to the emerging fields of data science and AI. These elements draw from the National Academies Roundtable on Data Science Postsecondary Education report (NASEM, 2018; NASEM, 2020) and the ASA-endorsed curriculum guidelines for undergraduate data science programs (De Veaux et al., 2017) and reflect nationwide discussions about how to develop a workforce that can harness the data revolution for all disciplines and professions. The ten elements are grouped into three categories and enumerated below:

**Data Perspectives**

1. Recognizing data as information – not truth – with error, variability and missing information.
2. Explaining what it means to be a data scientist and AI expert.
3. Observing a variety of data scientist role models and careers.

**Data Practices**

4. Examining how data are created, and the related assumptions and collection practices.
5. Practicing data curation, wrangling and cleaning.
6. Assessing validity of data, methods, results and communication.
7. Employing design practices such as documenting work, considering whether broadband is required for applications, including color palettes that are visible to people who are color blind, adding captions to video and adding descriptive text to images.
8. Investigating ethical issues and ways to approach them.

**Data Discoveries**

9. Articulating current issues or open questions in data science and AI.
10. Sharing exciting discovery or impact of data science and AI.

Instructors have flexibility in how these learning elements are embedded and highlighted in their courses. As part of the DSA instructional support structure, and for related ADAPT model implementation guidance, instructors attend topical DSA professional learning community discussions at least twice per month (DuFour, 2004; Stoll, et al., 2006). During these discussions, instructors may consider and address implementations of the 10 Common Learning Elements, including how they have brought them into the learning process, how they chose to engage students using the concepts, and challenges and ways that worked well. Discussions around the 10 Common Learning Elements often facilitate reflective understanding of the ways professionals practice data science and what it means to participate in data science from different disciplinary perspectives for both students and instructors. At times, these discussions extend to smaller interest groups beyond the general DSA space.

As an example of the evolutionary intentions of the ADAPT model, the 10 Common Learning Elements have been updated over time to reflect evolving trends in Data Science in AI based on stakeholder feedback.. An upcoming study will examine the variety of ways that the elements are implemented across different instructors and course topics, as an addition to examining model enactment.

**ADAPT Model Component 3: Workforce Preparedness**

The goal of the Workforce Preparedness component is to support students as they develop and sustain their identity as people capable of engaging in data science and AI (Carlone & Johnson, 2007). A primary way that this component appears in courses is through facilitating students' decision making opportunities and allowing them to build their data science and AI agency (e.g., choosing project datasets or analysis platforms and methods). Offering students a range of choices of both data contexts and data science methods allows students to explore career pathways and shape robust data science identities (Lesser, 2007; Weiland, 2017) while accumulating relevant statistical and technological skills (e.g., Çetinkaya-Rundel & Ellison, 2021).

With **data context choices**, students may choose to explore phenomena that are consequential for specific groups (e.g., exploring lack of broadband access by students from rural areas, Sanei, Griffith, & Jones, 2022). Instructors balance the benefits of encouraging individual student choice of data context with practical considerations. For example, instructors need to ensure that appropriate and quality datasets are available and that locating them is manageable within course time constraints.

With **choices of data science methods**, students can decide between different tools and concepts to engage in the data investigation process. Examples include choice of programming language, algorithm, visualization tool, or data modeling methodology (statistical, geographic, mathematical, etc). Students may also consider which practices they emphasize, for example, in communicating project insights.

The two categories of choices – context and methods – can be simultaneously embedded in student course and project selections to reflect students' interests and career goals. In the courses that are organized by topics or tools (e.g., Introduction to AI Ethics and Introduction to R/Python for Data Science), students may choose to focus on a tool that best fits the potential needs in their future applications of data science. Together, these two kinds of choices create flexible pathways into data science that meet the needs of students from majors in every department and college of the university.

Below is a quoted example of one student's reflection on their DSA courses experience directly related to the Workforce Preparedness emphasis of the ADAPT model:

> *"The Data Science Academy was instrumental in preparing me for the workforce. Through hands-on projects covering every stage of the project life cycle, I gained programming*

*experience and practical insight specific to my field of interest. Nontechnical skills such as project scoping, developing proof-of-concept frameworks, and documenting the data cycle aren't taught in traditional classrooms but are critical for preparing students for real-world data science projects. Those skills set me apart from my peers during interviews and made the transition into the workforce seamless."*

An additional support space for students and instructors alike involves the DSA Course Collaboration Leaders (CCLs). This group of undergraduate leaders who have taken DSA courses provide peer support to students in DSA courses and more general support to the DSA teaching and learning space through valuable student insights and innovation. The CCLs facilitate their peers' data science and AI agency with the leadership guidance they provide.

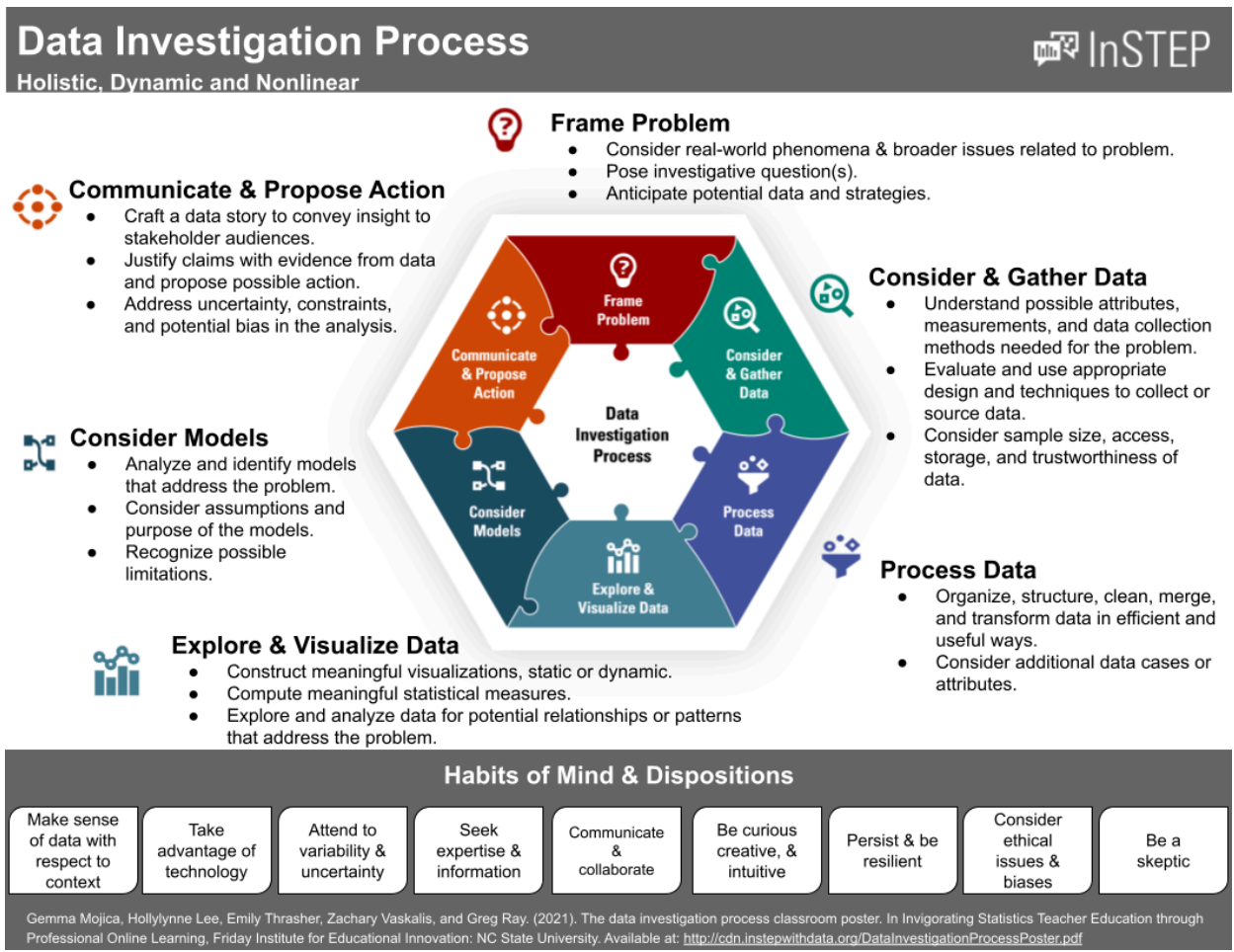

*Figure 3: Data Science Investigative Framework. Based on Lee, H., Mojica, G., Thrasher, E., & Baumgartner, P. (2022). Investigating data like a data scientist: Key practices and processes. Statistics Education Research Journal, 21(2), 3-3. Reprinted with permission.*

## A complementary data science model

In addition to the ADAPT model, the depiction below shows a data investigation framework that was developed in collaboration with practicing data scientists at Research Triangle International. The framework brings workforce relevant practices into school instruction by presenting a conceptual framework for how data scientists might go about accomplishing their work. This framework, which is referenced in some DSA courses, is also the basis for the most recent revision of North Carolina's CS 30: Introduction to Data Science course for middle and high school students. Note parallels to process cycles such as mathematical modeling as described in the Guidelines for Assessment and Instruction in Mathematical Modeling Education (Garfunkel & Montgomery, 2019), which can be extended to data science and AI development in learning and work contexts.

## Learning Pathways: ADAPT Courses and Credentials

A primary goal of the DSA 1-credit courses is to provide building blocks for learning around campus as a mechanism for institutional transformation. DSA courses provide new opportunities for students in all majors to explore data science and AI. The figure below shows some DSA curricular features with related descriptions.

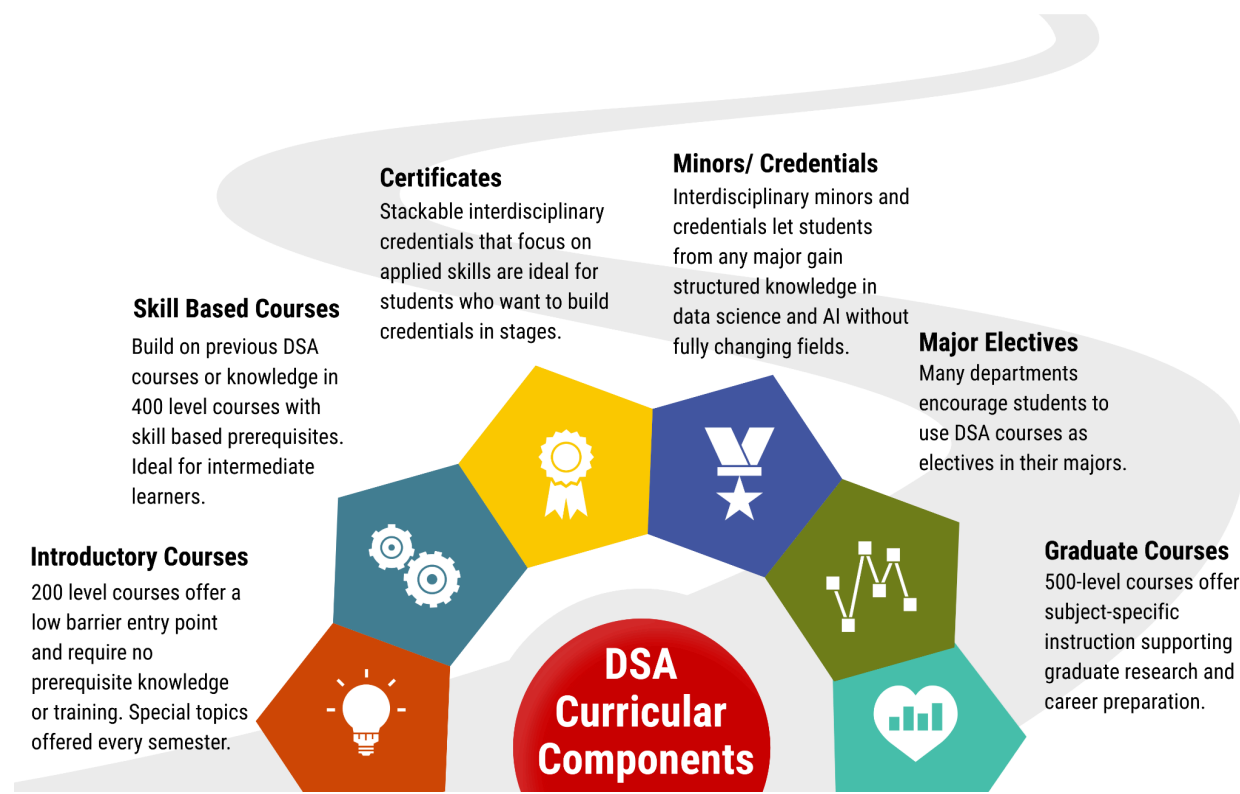

*Figure 4: Curricular components at NC State that incorporate DSA courses, including introductory courses, skill-based courses, certificates, credentials, minors, electives for majors and graduate courses (credit: Sarah Whichello).*

**Examples of DSA ADAPT Course Topics**

Interdisciplinary course offerings at multiple levels are offered each semester. In addition to permanent courses, special topics at multiple levels are offered as well. Below is a list of DSA course offerings from Spring 2026 as an example of how interdisciplinary DSA courses are featured within a semester.

- DSA 201 - Introduction to R/Python for Data Science ( 5 sections)
- DSA 202 - Introduction to Data Visualization (3 sections)
- DSA 205 - Data Communication
- DSA 220 - Introduction to AI Ethics
- DSA 225 - Data Science for Social Good
- DSA 235 - Introduction to Data Science for Cybersecurity
- DSA 240 - Measuring Success
- DSA 295 - Introductory Special Topics in Data Science (4 topics)
    - o   Virtual Reality Exercise & Personal Health Analytics
    - o   Citizen Science Data Analytics
    - o   Introduction to Network Analysis
    - o   AI for Data Science: A No-Code Introduction
- DSA 405 - Data Wrangling and Web Scraping
- DSA 406 - Exploratory Data Analysis for Big Data (2 sections)
- DSA 410 - Data Internship Preparation for Social Impact
- DSA 412 - Exploring Machine Learning (2 sections)
- DSA 435 - Predictive Analytics for Improving Services
- DSA 440 - Introduction to APACHE Spark Using Big Datasets
- DSA 495 - Special Topics in Data Science (3 topics)
    - o   Advanced Supervised ML Using a Visual Interface
    - o   Sports Analytics & Forecasting Using R
    - o   Algorithmic Fairness and AI Accountability
- DSA 595 - Graduate Special Topics in Data Science (1 topic)
    - o   Bayesian Computations for Machine Learning

**New credentials: minors and certificates**

Beyond DSA curricular components, in various collections, DSA courses can (and do) serve more broadly as a basis for *data science and AI pathways*. The credentials listed below were developed for this purpose in coordination with creative collaborators in existing NC State departments.

- Data Science in Business Minor
- Data Science with Graphic and Experience Design Minor
- Data Science in Engineering Analytics and Decision-Making Minor
- Mathematical Data Science Minor
- Data Science in K-12 Education Minor
- Interdisciplinary Applied Data Science Minor

In addition to the minors listed above, two 12-credit DSA certificates were developed. The Data Science in Business certificate is available to non-degree studies (NDS) students and extends data science and AI learning opportunities beyond full time NC State students. Currently, the Data Science with Graphic and Experience Design Certificate is restricted to students in the design major only, but has the capacity to accommodate NDS students in the future, as well.

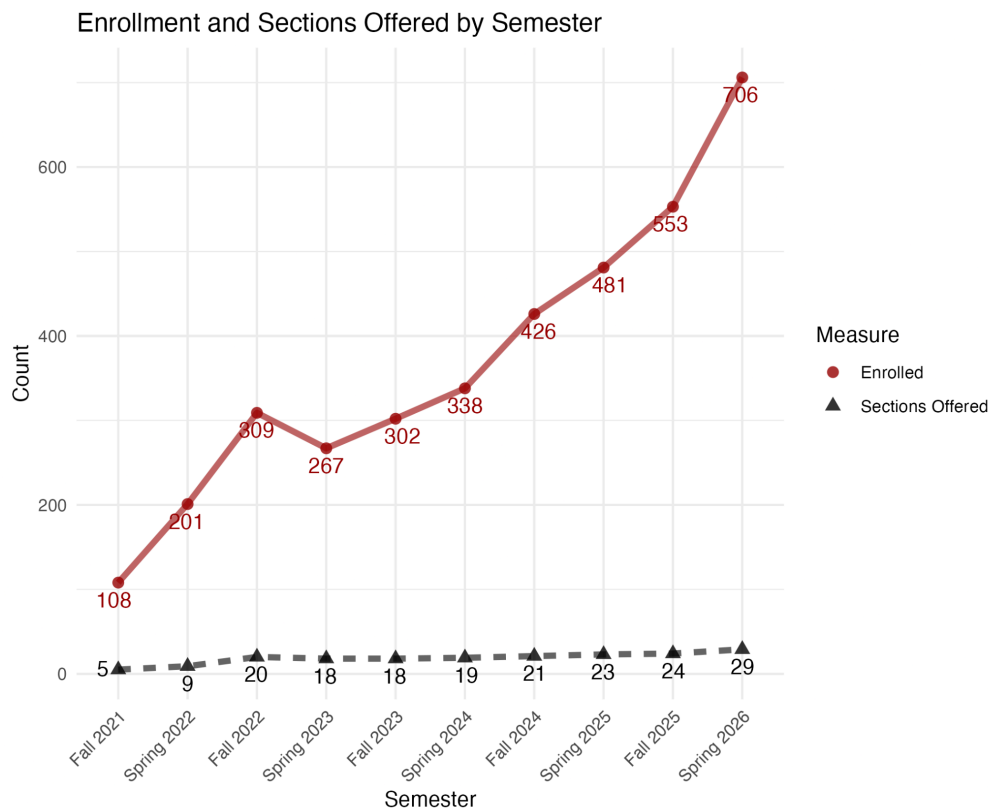

Figure 5: *DSA courses enrollment and number of sections offered over time, from Fall 2021 to Spring 2026.*

**How did we get here: A historical overview and current characterization**

In Fall 2021, the DSA curriculum began with five sections. In Spring 2026, the number of sections had expanded to 29. Enrollment had increased at a faster rate than section expansion and the average number of students per section reached the 25 seat cap. Currently, creative ways to scale up within the project-based ADAPT model structure have become a new and exciting challenge.

The section expansion and enrollment increases coincided with collaborative developments (articulations) that saw DSA courses included into existing programs and the development of DSA credentials. The first example of articulation, in 2022, occurred when faculty in the College of Textiles voted to accept up to three DSA courses as general electives for many of their majors. A more recent example includes the Certificate in Professional Writing, which incorporates 10 unique DSA courses (e.g., DSA 202 Introduction to Data Visualization, DSA 205 Data Communication, and DSA 220 Introduction to AI Ethics).

Currently, departments in all 12 NC State colleges (including the graduate school) have articulated courses into majors, minors, concentrations or certificates. This is reflected in the variety of majors (approximately 188) represented by students in DSA courses within the first 5 years of offerings (see Figure 4).

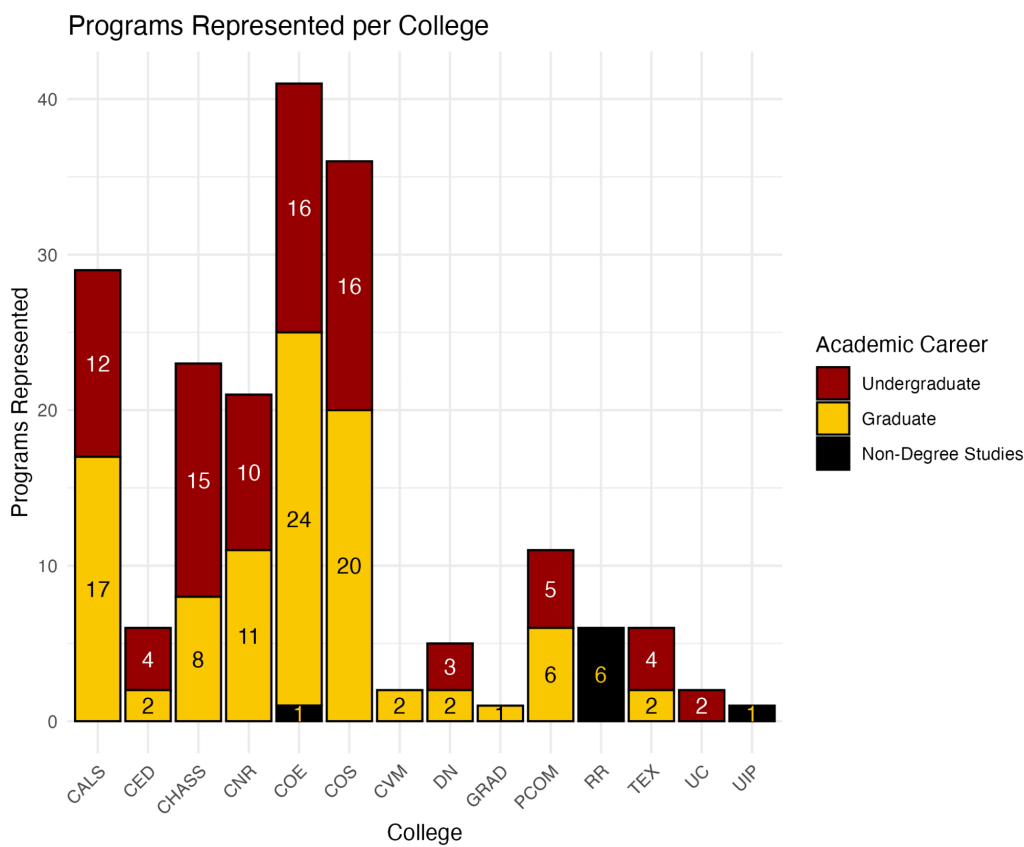

*Figure 6: Number of majors represented in DSA Courses Per College, categorized by undergraduate, graduate, and non-degree studies programs. The 188 distinct majors/programs include representation from each college. Included in this count are two unique "Engineering First*

*Year" representations from COE and CNR, and two unique "Life Sciences First Year" representations from COS and CALS for a total of 190. . Appendix A contains a key to the college name acronyms and the total enrollment of each college.*

Figure 7 below shows how the number of majors represented by a particular college varies and is not necessarily strongly associated with the number of credits taken by students within a particular college. Beyond the number of majors a particular college offers, to some extent this trend is related to data science and AI credential pathways, where there may be a need for development not only within certain colleges, but also in departments within a given college. Data science and AI pathways already exist within many departments and this presents an opportunity for potential collaborations and distinguished developments.

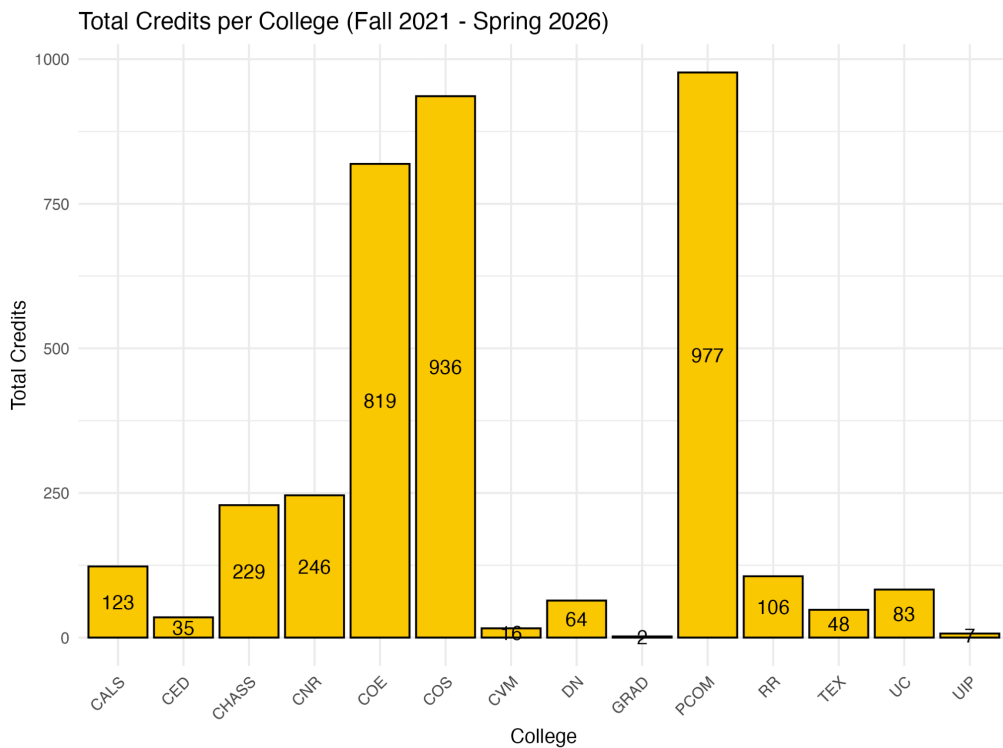

*Figure 7: Total Credits Per NC State College., Although the credits breakdown does not directly mirror student population percentages across colleges, participation has broadened steadily to include students from every college and many of our 86 departments. Appendix A contains a key to the college name acronyms and the total enrollment of each college.*

Figure 6, below, yields additional insight into the variety of students in DSA courses categorized by academic career. The breakdown largely reflects the DSA focus on undergraduate courses and pathways.

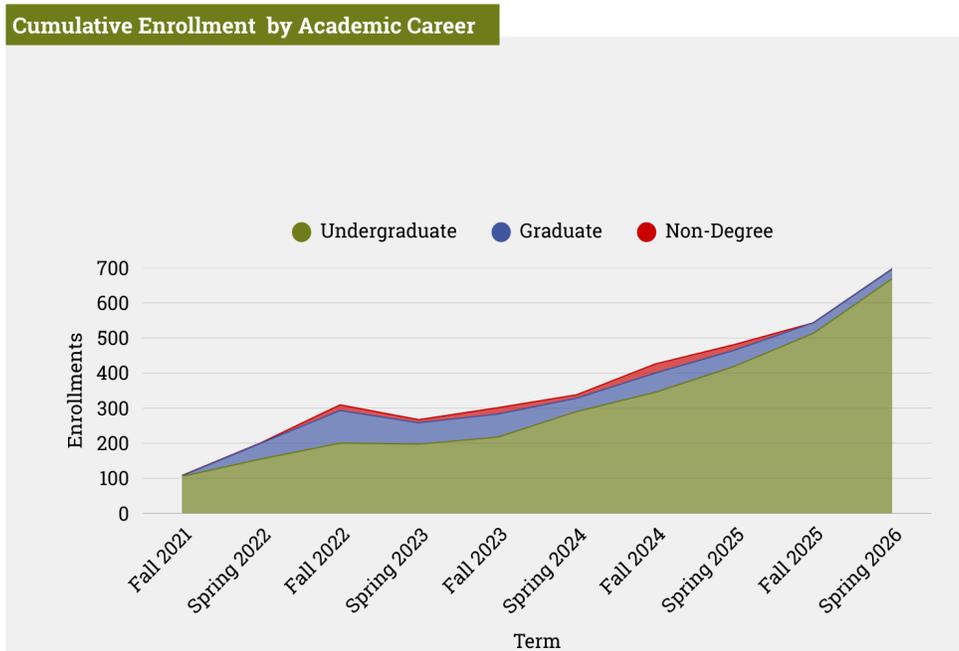

*Figure 8: Cumulative Enrollment by Academic Career. Enrollment and number of sections offered each semester. Note that enrollment dipped in Spring 2023 when the number of sections was temporarily reduced. Notably, enrollments include faculty, staff and community members, some of whom are enrolled in degree programs or certificates, and others who have joined as non-degree studies students. Appendix A contains a key to the college name acronyms and the total enrollment of each college.*

**Data Science and AI Education Research and Adoption**

A fortunate consequence of developing the ADAPT model from the beginning of the DSA and also developing an IRB-approved research study in which we could consent students for their permission to collect their work is that we have been awarded two National Science Foundation grants to study the teaching and learning in the DSA using the ADAPT model, an Innovation in Undergraduate STEM Education award DUE-2313644 and a Postdoctoral Cohort DGE-2222148. The projects supported by these awards have involved an interdisciplinary group of researchers, largely led by faculty in the NC State College of Education and involving graduate students, postdoctoral researchers, DSA students and instructors in DSA and 2- and 4 year partner institutions.

Researchers involved in ADAPT research projects have included faculty Dr. Rachel Levy (PI), Dr. Sunghwan Byin (co-PI), Dr. Shiyan Jiang (co-PI), and Dr. Ela Castellanos-Reyes (co-PI); postdocs Dr. James Harr, Dr. Zarifa Zakaria, Dr. Jeanne McClure, Dr. Kelsey Dufresne, Dr. Tom Leppard; and graduate students, Alin Yalcinkaya, Nixon Igunza, and Michelle Pace, Doreen Mushi, Hamid Sanei, Matthew Ferrell, Laurie Short, Ismail Hossain and Bouchra Elgaou. In addition, Dr. James Harr (postdoc) and Nick Kruskamp (graduate student) were involved in early development of the

ADAPT model. Dr. David Stokes and Mr. Mahmoud Harding produced a book based on the ADAPT Introduction to R/Python courses, and Dr. Tom Leppard produced a book based on the ADAPT Network Analysis Course.  This work was supported by the NC State Provost's office and two National Science Foundation grants (*#DUE: 2313644 and #DGE-2222148*).

We encourage you to follow these researchers for other publications and products related to the ADAPT model. In some cases, even just reading about the model has inspired curricular innovation, as described below:

> *NC State's Data Science Academy's ADAPT model is so well developed and tested that it has proven invaluable to our own emerging curriculum development around data science and statistical pitfalls. We became interested in ADAPT as it integrates technical skills within the broader context: students solve real data problems and learn to exercise choices regarding tools or research outcomes, a key skill needed in jobs where often neither are specified. The model's framework that makes it clear that information can be biased, with errors, and not be representative, also fits well with our statistical pitfalls course. Thank you for making ADAPT openly accessible!*
> *-- Jay Cordes and Julia Koschinsky (https://puttingscienceintodatascience.org)*

Instructors have also described the ways that teaching DSA courses has positively influenced their impact in their own careers:

> *"Teaching in the DSA has taught me to expand my world view a bit, and to be more sensitive and more aware of where my paradigm doesn't fit the needs of others.  While I highly value consistency in my daily work practices, I've gone to appreciate the need to be more flexible and, as I tell my students, to meet my audience where they are.  DSA has given me a great opportunity to learn from others' paradigms as well as to refine my own, and I'm grateful to you for that."*
> *– DSA Instructor Barbara Prillaman*

**ADAPT Model Courses in Industry and Government**

In addition to courses primarily focused on university learners, the DSA has developed workshops and courses that are non-credit-bearing. These courses are collectively called **Data and AI at Work**, to signal that the courses will be customized to the data and problems that the learners will encounter in their workplace.  As an example, the North Carolina Department of Health and Human Services offers a three course series to their employees along with a train-the-trainer model in which employees who start as learners can work toward being able to deliver future training. These courses all follow the ADAPT course model and are entirely project-based. The DHHS sequence is Advanced Excel, Power BI and a capstone project, whereas other clients have requested instruction using JMP, SAS, R and Python.

Some other examples include a new collaboration with Ablr will co-develop workforce-centered AI literacy courses for individuals with low vision or blindness. A series of workshops in collaboration with North Carolina Community Colleges will bring the ADAPT model of teaching and learning to cohorts of instructors who are in leadership roles in community colleges across the state. And the ADAPT model is central to our work in K-12, both modules for students and teacher professional development. In particular, the revision of CS 30: Introduction to Data Science, which is possible to offer in middle and high school and satisfies the high school computer science graduate requirement was developed with the ADAPT model in mind.

**ADAPT in the Community College Setting**

In Spring 2026, NSF-IUSE collaborators Levy, Byun and Castellanos-Reyes partnered with Andrea Crowley, Lane Freeman and Dana Newton to present two ADAPT model workshops to community college master instructors across North Carolina. This set of workshops was a pilot to see how instructors might develop project-based lessons that use the ADAPT model in a community college setting across disciplines as wide ranging as healthcare, humanities, STEM, HVAC and emergency management.  Figure 8 shows the reach across over half of the 100 counties of North Carolina in only two in-person workshops.  In the workshops the master instructors were introduced to the ADAPT model and basic principles of data literacy, discussed the types of data used in their disciplines, identified which of the 10 learning elements they would like to explore with their students, practiced using no-code data exploration tools such as CODAP from Concord Consortium, and planned a way to incorporate a data investigation into one of their courses.

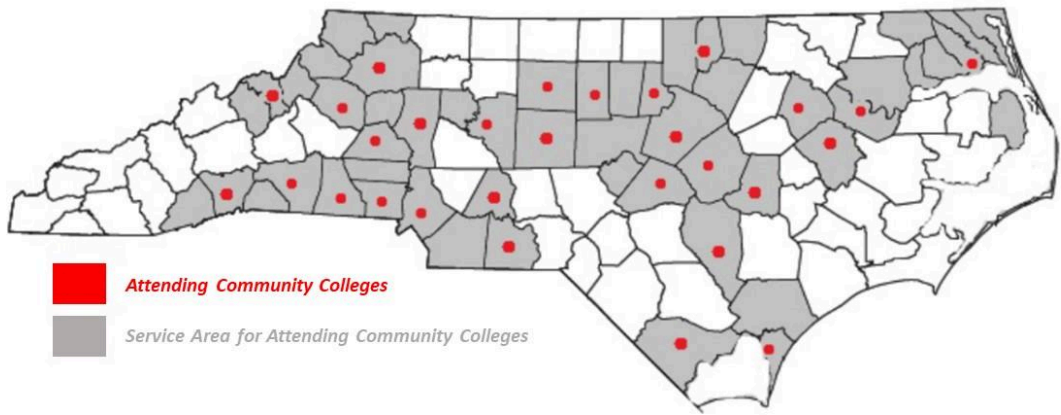

Figure 9:  Community Colleges represented by Master Instructors attending two Spring 2026 ADAPT workshops (red dots) and the 54 unique NC counties in their community college service areas (shaded counties).  Excerpted from Figure 3 of external evaluation from Hasse, Lovin and Associates.

After the workshop the participants completed a survey to indicate their overall level of satisfaction with the workshop, their experience of the various components of the workshop and how they intend to use and communicate with peers about the ADAPT model and data investigations going forward. The participants used a 5-point scale (*1-Very Dissatisfied* to

*5-Very Satisfied*) to rate their overall level of satisfaction with the workshop. Figure 9 contains the averages of the responses from the two workshops and overall. Participants rated their satisfaction between *Satisfied* and *Very Satisfied*.

| Survey Question | Feb 6 (n=21) | Feb 20 (n=30) | Overall (n=51) |
|---|---|---|---|
| How would you rate your overall level of satisfaction with the workshop? | 4.67 | 4.57 | 4.61 |
| This workshop was informative. | 4.76 | 4.80 | 4.78 |
| This workshop was engaging. | 4.86 | 4.77 | 4.80 |
| This workshop was worth the time invested. | 4.86 | 4.77 | 4.80 |
| Information from this workshop can be used in my instruction. | 4.81 | 4.70 | 4.75 |
| I am more confident to integrate the presented concepts in my instruction/work. | 4.52 | 4.47 | 4.49 |
| I would recommend this presentation to other community college instructors/staff. | 4.81 | 4.63 | 4.71 |
| I would be interested in integrating the ADAPT module into my instruction. | 4.71 | 4.67 | 4.69 |
| I would be interested in additional data science/AI workshops. | 4.62 | 4.80 | 4.73 |

Figure 10. Table of evaluation data from external evaluation by Hasse, Lovin and Associates of the Community College workshops. The participants used a 5-point scale (*1-Strongly Disagree* to *5-Strongly Agree*) to rate their level of agreement with statements. Participants rated each statement between *Agree* and *Strongly Agree*.

*We were able to make adjustments between the workshops* based on participant feedback such as providing resources and links before the workshop to allow participants to become familiar with content, continuing to curate data science examples from diverse areas of study, such as English and health sciences, grouping participants by discipline to encourage collaboration and providing more time for practice with data science tools.In order to increase the range of the workshop, one participant suggested to reach more people this could be an online asynchronous training. Overall we were very encouraged by the ideas and conversation within the workshop and the followup email contact by several of the participants. We look forward to scaling up the delivery of these workshops including building capacity for people to facilitate the workshops on their own to reach more instructors and counties across the state.

**How to learn more about ADAPT and get involved**

The DSA website will be the most up-to-date resource on current courses and materials to guide instruction, opportunities to join events and how to reach us by email (you can use our general email for all inquiries).  We welcome visitors to our professional learning community which meets online.  We look forward to hearing from you about how the ADAPT model relates to your work.

We have related publications including two new books: the R/Python book by Stokes/Harding can be accessed by searching online and is being revised in response to feedback – we welcome yours!  In the references you will find links to the R/Python book, a network analysis book and ADAPT related research articles. DSA has a number of ways that we provide workshops to external groups through **Data and AI at work**.

We have an annual **ADAPT ShareFair** in which instructors and students share their experiences with the ADAPT mode and provide examples of student work. You are welcome to join and can find that information on the DSA website. We'll leave you with a quote from a student presenting at that event.

> *I come from a community college, so I was thinking about my engineering and design project in my first year at Craven Community College. The part that my team could not figure out was the electrical part.*
>
> *So at the time, I was thinking about doing chemical engineering. But that project changed my major. I was like, "Oh, I don't know how to wire this machine. How do I put a potentiometer so that my bubble machine can change and vary speed?"*
>
> *And that was the moment that I decided, because this is the part that I struggled with, and I could not complete for my design project. I was the project manager at the time. So I was thinking, I want to do electrical engineering.*
>
> *And that's how it brought me to NC State, and then brought me to the Data Science and AI Academy. And you know what — just starting in 2025, one of the major concentrations in electrical engineering is data science and AI.*
>
> *So everything connected every dot. Everything happened for a reason.*
>
> <div align="right">Vy Tran (NCSU, Class of 2028)</div>

**Acknowledgements**

The authors of this article are listed alphabetically as authorship was a collaboration with both contributing to the inception and execution of this paper. Dr. Rachel Levy, Executive Director of the Data Science Academy and Professor of Mathematics at North Carolina State University, initiated this report and was the lead on the development of the ADAPT model and writing. Dr. James B. Harr III, Assistant Teaching Professor of Data Science at the College of William and Mary and Instructor of Data Science at North Carolina State University, was the Postdoctoral Teaching Scholar and Teaching Coordinator when the DSA was founded and the instructor of record for the three case studies included in the article. Dr. David Stokes is the current Director of Data Science and AI Academic Programs, has supervised the development of courses and credentials and conducted the analysis of the course registrations.

The authors would first like to thank Dr. Sunghwan Byun, Dr. Shiyan Jiang and Dr. Ela Castellanos-Reyes, who have joined DSA education research as collaborators. We would like to thank Dr. Alyson Wilson and Dr. Emily Griffith whose early efforts in proposing the concept of a Data Science Academy and 1-credit courses have proven to be abundantly successful. We would like to thank Sarah Whichello for data visualizations and graphic design. Finally, we would like to thank the entire DSA Staff and affiliated faculty, our students and instructors, graduate and postdoctoral researchers, the NC State University Provost's Office, the Office of University Interdisciplinary Programs and the Office of Research and Innovation.

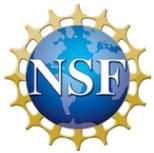

*This material is based upon work supported by the National Science Foundation under Grant #DUE: 2313644 and #DGE-2222148.  Any opinions, findings, and conclusions or recommendations expressed in this material are those of the author(s) and do not necessarily reflect the views of the National Science Foundation.*

**Appendix A**

| NC State College | Total Enrollments |
|---|---|
| College of Agriculture and Life Sciences (CALS) | 4,501 |
| College of Design (COD) | 955 |
| College of Education (CED) | 1,721 |
| College of Engineering (COE) | 12,111 |
| College of Humanities and Social Sciences (CHASS) | 4,881 |
| College of Natural Resources (CNR) | 2,138 |
| College of Sciences (COS) | 4,333 |
| Wilson College of Textiles (TEX) | 1,104 |
| College of Veterinary Medicine (CVM) (Graduate Only) | 604 |
| Poole College of Management (PCOM) | 4,734 |
| University College (UC) (Undergraduate Only) | 1,345 |

*Total student enrollment, undergraduate and graduate, by NC State College. Note that in addition to the NC State colleges above, Figure 6 and 7 include numbers from units categorized as RR (Registration and Records) and UIP (University Interdisciplinary Programs).*